\documentclass[12pt,a4paper]{article}

\usepackage{amsmath}
\usepackage{amsthm}
\usepackage{amscd}
\usepackage{amsxtra}
\usepackage{upref}
\usepackage{amsfonts}
\usepackage{amssymb}
\usepackage{graphicx}
\usepackage{url}
\usepackage{cite}
\usepackage{pifont}

\def\beq{\begin{equation}}
\def\eeq{\end{equation}}
\def\bea{\begin{eqnarray}}
\def\eea{\end{eqnarray}}

\theoremstyle{definition}

\numberwithin{equation}{section}

\title{Localization on the Landscape and Eternal Inflation.}
\author{Laura Mersini-Houghton$^{a,b}$ and Malcolm J. Perry$^a$\\
Department of Applied Mathematics and Theoretical Physics,\\
University of Cambridge,\\
Wilberforce Road,\\
Cambridge CB3 0WA,\\ England$^{a}$,\\Department of Physics and Astronomy, UNC Chapel Hill, NC 27599, USA$^{b}$.}
\begin{document}
\maketitle
\date{\today}
\begin{abstract}
We investigate the validity of the assertion that eternal inflation populates the landscape of string theory. We verify that bubble solutions do not satisfy the Klein Gordon equation for the landscape potential. Solutions to the landscape potential within the formalism of quantum cosmology are Anderson localized wavefunctions. Those are inconsistent with inflating bubble solutions. The physical reasons behind the failure of a relation between eternal inflation and the landscape are rooted in quantum phenomena such as interference between wavefunction concentrated around the various vacua in the landscape.
\end{abstract}
\section{Introduction}
Fundamental questions in cosmology require a proper quantum theory of gravity. The reasons are twofold:
the universe was sufficiently microscopic at its earliest moments that it needs to be considered as a quantum system and also that gravity played a central role in its existence from its beginning up till
the  present time. Despite the absence of a fundamental theory of quantum gravity, understanding the
origins of the universe and the global structure of spacetime, is not hopeless. String theory is the leading candidate at present for the theory of quantum gravity. Whilst string theory is not yet a complete description of quantum gravity, it has provided us with a picture of its vacua, coined the landscape. Somehow, the landscape should give rise to universes like ours. We also have the formalism of canonical quantum gravity, often called quantum cosmology, by which we can attempt to calculate and derive answers to fundamental questions such as the origins the universe. Quantum cosmology  applies quantum mechanics to the whole universe. It is a quantum version of general relativity that addresses issues related to the emergence of our universe and the global structure of spacetime. In the Hamiltonian formalism of quantum cosmology, the classical spacetime itself is not present from the start, it rather emerges later on from the evolution of the 3-metrics and their conjugate momenta promoted to quantum operators. The constraints based on Einstein equations become quantum operators ~\cite{kieferqm, kiefer, isham}. 
  
In this work we apply some ideas of quantum cosmology to the landscape of string theory with the aim of analyzing the validity of the claim made in ~\cite{susskind} that the string theory landscape is populated by  bubbles of expanding universes characteristic of eternal inflation. These bubble universes
live in different vacua of the landscape and thus are expected to be characterized by different low energy physics in different bubbles.
Considering the prominent role that the inflationary paradigm plays in cosmology 
the possibility of populating the landscape in this way needs to be taken seriously ~\cite{followers}, and then requires a rigorous justification. The finding of  wavefunctions in which our
universe comes out of the landscape then  becomes an imperative. 

\subsection{Review of the Claim for \lq Eternal Inflation Populating the Landscape'}
The proposition that eternal inflation populates the landscape \cite{susskind} is based on simplifying the dynamics of a quantum field propagating on an N-vacua landscape by that of a double well potential yielding solutions that are approximately Coleman-DeLuccia bubbles.   
In order to check whether this approximation is correct, we first review the arguments made in ~\cite{susskind} for inferring that the landscape is populated by eternal inflation bubbles. 

As is well known the landscape contains about $10^{600}$ different vacua \cite{landscape}. In \cite{susskind} it is assumed that it suffices to approximate this complicated structure with an asymmetric double well potential, the shape and slope of which have been chosen in a very special manner. The main argument given to reduce the dynamics on the vast landscape to that of a double well potential is based on appealing to the nearest neighbor approximation for interactions among vacua of the landscape. It seems reasonable to assume that our vacuum would be influenced only by its nearest neighbor vacuum, and thus conclude that the relevant portion to our vacuum is a double well corner on the landscape. Accordingly, if our universe is initially stuck in the false vacuum, it can only tunnel from its false vacuum state to its nearest neighbor that is then the true vacuum state. The latter must then be an anthropically arranged  double well reduction of the landscape potential. The shape of the double well potential is very special:  the neighbor to our initial vacuum must be the true vacuum; the potential must have a slope flat enough to allow for inflation in the bubble after it tunnels through to the true vacuum;  the flat slope of the potential should be long enough to allow for at least $60$ efoldings to occur as the field rolls down towards the true vacuum neighbor. When a sufficient number of efoldings have occurred, then the slope of the potential around the true vacuum must become suddenly very steep so that reheating  occurs and finally the energy in the true vacuum neighbor should be about $122$ orders of magnitude less than the Planck energy in order to produce the observed dark energy present in the universe. The extraordinarily fine tuning of this potential \cite{susskind} is due to the anthropic selection. Our true vacuum neighbor must also have been carefully selected to produce an open universe bubble that resembles closely our universe, including inflation by the right number of efoldings and reheating epochs. In this scenario, false and true vacua are separated by a high and narrow barrier, to guarantee the bubble's thin wall approximation. 

Another piece of anthropic tuning of the potential comes from the fact that bubbles contain open universes with curvature $k=-1$ in their interior. If our universe is contained in the interior of the bubble, then the curvature contribution to the energy density of the universe goes as $1/a^2$ (where $a(t)$ is the scale factor). Since curvature dilutes much more slowly than radiation ($1/a^4$) and matter ($1/a^3$) in the universe, then it must be tuned by $60$ orders of magnitude to be exquisitely small at nucleation in order to agree with present  observational bounds of $\Omega_k \le 0.02$. The curvature is related to the shape of the potential $V(\phi)$: the condition for tunneling and bubble nucleation is given by $V'' \gg V$ and the condition for the slow roll inflation after nucleation requires $V'' \ll V$ . Therefore the part of the potential between bubble nucleation and the slow roll regime that contributes to the curvature of the universe has to be exquisitely fine tuned. The purpose and usefulness of the vast and complex landscape with $10^{600}$ vacua in this scenario is found in simply justifying anthropically the 'pure chance' of finding a highly contrived double well potential. The existence of the landscape becomes useful in increasing the chances of finding a corner of it that looks like the fine-tuned double well potential described. If such a corner exists then it will be 'lifted' out of the landscape anthropically.

Using the landscape for simply justifying the choice of an unusually shaped double well potential as the one described above, is the least worrying aspect of this claim. An implication of this scenario is that since our patch can only tunnel to the nearest neighbor vacuum, it therefore has no correlations with the rest of the landscape, then the remaining  $(10^{600} - 2)$ vacua on the landscape become completely irrelevant. If that were the case, then the existence of the landscape becomes unfalsifable.  We show in Section $6$ and discuss in the last section, our concern with this scenario arises when all the physical phenomena that stem from the intricate and complicated structure of the landscape, such as: interference, diffusion and multiple scattering leading to transport of the field through the N-vacua of the landscape, are ignored. Quantum interference is always present in an N-vacua landscape, alongside tunneling to the nearest neighbor. These two effects play a crucial role in transporting the field over many vacua of the landscape even in the nearest neighbor approximation. These are some of the reasons why the double well bubble solution turns out to be inconsistent with the landscape potential. In reality, the quantum dynamics of the field on the rich landscape gives rise to completely different field dynamics and highly nontrivial solutions ~\cite{laura} from those found in eternal inflation scenarios. We conclude that the dynamics of interference and multiple scattering of the wavefunctional of the universe on the landscape, are best described by Anderson localized wavepackets. 

We mentioned the problems that arise by oversimplifying the landscape potential to that of a double well potential. A new family of problems, including mathematical and conceptual inconsistencies, arise when investigating this claim. Perhaps however, the most mysterious problem is that the field fluctuation is a random variable. Therefore the background geometry for bubbles embedded in the false vacuum de Sitter space is a random variable. The metric of each bubble is also an independent variable. On these grounds, the study of eternal inflation on the landscape should allow the geometries of false and true vacuum to vary randomly on the landscape. But a field theory approach of populating the landscape with eternal inflation bubbles with a classical apriori fixed background, is inconsistent with the expected randomness of fields and geometries on the landscape. The interpretation of eternal inflation populating the landscape is equally difficult to achieve due to the lack of a formalism where not only the field but also all background and interior geometries of the bubbles are allowed to be independent variables. Further, the fact that eternal inflation is based on field diffusion implies that the field will not simply visit the minima of the idealized double well, it will also diffuse to other vacua in the landscape beyond the double well corner, and so the double well idealization will break down. This motivates us to use quantum cosmology in the study of field solutions on the landscape, since it allows the emerging geometries and field configurations to be treated as independent quantum variables.

\subsection{Mathematical validity of Eternal Inflation Bubbles on the Landscape ? }

Before we embark upon the formalism of quantum cosmology in Section $3$ and its application to the landscape, we would like to first check whether the Klein Gordon equation for a landscape potential, is satisfied by solutions resembling eternal inflation bubbles as postulated in ~\cite{susskind}.
The Klein-Gordon equation for a field $\phi$ in a potential $V(\phi)$ is

\begin{equation}
\Box \phi  + V'(\phi)  =  0  \,
\label{KG}
\end{equation}

The solution for $\phi$ from this second order differential equation will depend on the choice of the potential $V(\phi)$. The solution for the case of a double well potential $V_{t}(\phi)$ is the well known bounce solution $\phi_{t}$. Denote the landscape potential by $V_{L}(\phi)$, and its respective solution for the field by $\phi_{L}$.  We expect that different potentials in equation ~\ref{KG} will produce different solutions $\phi$. 
Let us test if bubble solutions $\phi_t$ approximately satisfy the Klein Gordon equation for the landscape potential $V_L$ in equation ~\ref{KG}. We can verify this claim by taking the bubble solution for $\phi_{t}$ resulting from false vacuum decay in the double well potential, with an $k=-1$ inflating interior, as posited in ~\cite{susskind}, and plugging it into the Klein-Gordon equation ~\ref{KG} with a potential $V = V_L$. At the end we explore issues related to the conceptual interpretation of these solutions especially when gravity becomes important. 

\subsubsection{Review of the Landscape Potential}
In order to check the validity of bubble solutions to the landscape Klein-Gordon equation, we first need to describe briefly the distributions of vacua and the structure of the landscape potential $V_{L}(\phi)$. As shown in ~\cite{denef} the distribution of landscape vacua and their energies is quite close to a random white noise distribution. Each of the vacua has a large number of internal degrees of freedom $\xi_j$, depending on moduli and fluxes. However it was shown that the internal degrees of freedom $\xi_i$ for each vacuum $\lq i \rq$ are Gaussian peaked around some mean moduli field value $\phi_i$. The latter allows us to treat this modulus $\phi$ as a collective coordinate and therefore reduce the landscape potential $V(\phi)$ to an one dimensional potential of a scalar field $\phi$. As explained in ~\cite{laura}, even within the assumption of the nearest neighbor approximation, the landscape potential is captured by the following form

\begin{equation}
V_{L}(\phi) = V_{0}(\phi) + V_{I}(\phi)  \,
\label{landscapepot}
\end{equation}
where $V_{0}(\phi)$ is a function that draws random values from an interval $[0,W]$ for each vacuum, $\phi_i$ thus representing the value of the \lq unperturbed' vacuum energies $V_{0}(\phi_i) = \epsilon_i$ at each vacuum site $\phi=\phi_i$. In other words $V_{0}(\phi)$ is diagonal when evaluated with respect to the states and those diagonal terms are drawn randomly from the interval $[0,W]$. The 'interaction' term of the potential $V_{I}(\phi)$ describes the field's interaction that allows for transport of the field, including tunneling, among the various landscape vacua. Therefore $V_I$ will typically contain off-diagonal terms when evaluated with respect to the states $\phi_j$. In the nearest neighbor approximation for a potential with a random distribution of vacua with strength $W$,  we thus have

\begin{equation}
\langle V_{I}(\phi) \rangle = 0  \qquad  \langle V_{I}(\phi_i) V_{I}(\phi_j) \rangle = W \delta(\phi_i - \phi_j)  \,
\label{interactionpot}
\end{equation}
where $\langle...\rangle$ describes averaging with respect to the states ${\phi_j}$. In most cases, where $V(\phi)$ is a randomly valued function, the details of disorder do not matter. The key point for these potentials is that such \lq disordered lattices' give rise to transport of the field due to tunneling and interference of phases from multiple scattering among  many vacua sites. These two physical mechanisms of transport can connect the initial vacuum site with many others on the landscape and not simply to the nearest neighbor. The phase superposition from multiple scattering leads to constructive and destructive interference which ultimately produces localization of the wavefunction in some vacua $\phi_i$. The site $\phi_i$ where the wavefunction will localize can not be predicted apriori. For the case when disorder vanishes, $W \rightarrow 0$, the two mechanisms of tunneling and interference, lead to perfect transport of the wavefunction. These (delocalized) solutions are known as the $\theta$-vacua solutions,  or Bloch waves. The latter class of solutions, valid for the supersymmetric sector of the landscape ($W=0$), do not give rise to universe solutions since the wavefunction cannot be localized around a vacuum, therefore this sector is irrelevant to our discussion. Here we focus on the landscape sector with $W \ne 0$. Due to randomness, the relevant quantities are not the field solutions but their probability distribution. 

Without losing generality, let us consider, for illustration, the distribution of vacua energies to be a Gaussian 

\beq
P(\epsilon) = \frac{1}{\sqrt{\pi NW}} e^{ - \frac{(\epsilon - \tilde{V}_0)^2}{NW} }
\eeq
where $NW$ is the mean 
width of distribution and $\tilde{V}_{0}$ denotes the ensemble averaged mean value $\langle \phi_{1} | V_{0}(\phi) | \phi_{N} \rangle = \tilde{V}_{0} $ .

For $W$ large, of the order of the string scale,  the randomness of vacuum energies from site to site leads to a very wide distribution which does not really have any peak. For the case of large disorder then, the solution to this problem cannot be done perturbatively or by the Wigner-Dyson method. Instead, the problem is solved by methods of Random Matrix Theory (RMT) ~\cite{efetov,altland}, where $P(\epsilon)$ is promoted to a quantum operator on the space of randomly realized Hamiltonians $\hat{H}$, 

\begin{equation}
{\cal P}[ \hat{H}(\phi)]  \sim {\mathcal N} e^{ - \frac{( \hat{H}(\phi) - \tilde{V}_{0} )^{2}}{NW} } \, 
\label{rmtprobability}
\end{equation}
where ${\mathcal N}$ is some normalization constant.

Let's take $\tilde{V}_0 = 0 $ from now on for simplicity, since it just denotes an overall shift of all vacuum energies by that amount, which we can scale to zero. Equipped with the structure of the landscape potential $V_{L}$, we are now in a position to verify whether this potential can allow eternal inflation bubbles to populate its vacua.

\subsubsection{Bubbles on the Landscape}
It should be noted that due to disorder, translation invariance is not a symmetry of solutions on the landscape. Due to gravity, parity invariance is also broken since reflection to negative energies, $\epsilon \rightarrow -\epsilon$, is not allowed. In such a situation, it becomes clear that we need quantum cosmology for treating the problem since the \lq bubble' geometries one hopes to obtain are all causally disconnected from each other, therefore these geometries should be treated as a free variable. Furthermore, even the embedding background with a false vacuum $V_{0}(\phi_j)$ and geometry described by a set of variables $a_j$, from which these \lq bubbles'  nucleate, is itself as we explained above, a randomly chosen variable drawn from an interval of energies $[0,W]$. Applying quantum cosmology to the landscape potential allows us to treat geometries of \lq bubbles' and the background as independent variables. Before proceeding with quantum cosmology where both the geometry and the field are independent variables in the minisuperpsace, let us in this section use the same approach as the one in ~\cite{susskind} and rely on quantum field theory and general relativity equal, since our aim is to verify their results. Despite some interpretational shortcomings ~\cite{kiefer, isham}, one can think of the quantum field theory treatment as the limit of quantum cosmology in the Born-Oppenheimer approximation.

Following the same setup and reasoning as the authors of ~\cite{susskind}, we set up the field at some initial vacuum $\phi_j$ which has been chosen anthropically to be of the \lq right' kind. Then the equation of motion for the field is

\begin{equation}
\ddot\phi + 3 H \dot\phi = - V_{L}'(\phi) \,
\label{fieldeom}
\end{equation}
 and the $00$ component of the Einstein equation provides the ensuing geometry of the field located at $\phi_j$ vacuum
\begin{equation}
\left(\frac{\dot a}{a} \right)^2 = \frac{8 \pi G}{3}\left(\frac{\dot\phi^2}{2} + V(\phi)\right) + \frac{1}{a^2} \,
\label{fieldgeometry}
\end{equation}

The last term comes from the curvature contribution of an open FRW universe ($k=-1$) contained in the interior of the \lq bubble' with metric
\begin{equation}
ds^2 = dt^2 - a(t)^2 \left[dr^2 + \sinh(r)^2 d\Omega^2 \right] \,
\label{bubblemetric}
\end{equation}

Initially the curvature term dominates thus $a(t)\simeq t$. Later on the bubble interior is supposed to undergo inflation, as the bubble slow rolls down the specially chosen flat portion $V(\phi_j)$ of the landscape, thus later $a(t) \simeq e^{Ht}$. Bubbles are $O(4)$ symmetric and when analytically continued to imaginary time $t \rightarrow i\tau$, the bubble becomes part of a 4-sphere in Euclidean space, described by the solution
\begin{equation}
\phi = \frac{\mu}{\sqrt\lambda} \tanh \left[\frac{\mu}{2}\left(\rho - \tilde\rho \right)\right] \,
\label{fieldsolution}
\end{equation}
where the mass term is related to the curvature of the potential around the barrier, $\mu^2 = V''(\phi_j)$, and $\rho$ is the Euclidean distance. Let us now use $ dot $ to denote the derivative with respect to Euclidean time $ d / d\tau$. The Euclidean equation of motion is ~\cite{cdl}
\begin{eqnarray}
\ddot\phi +3 H_{E} \dot\phi = V_{L}'(\phi) \, \\ \nonumber
H_{E}^2 = \frac{1}{\rho^2} + \frac{8 \pi G}{3}\left(\frac{\dot\phi^2}{2} - V_{L}(\phi)\right) \,.
\label{cdlequations}
\end{eqnarray}

From the above equations, the only difference between the bubble solution case and the landscape case, stands in the nature of the potential $V(\phi)$. In our case $V_{L}(\phi) = V(\phi)$ in the neighborhood of a specially chosen corner $\phi_j$ on the landscape is given by $V_{0}(\phi_j)=\epsilon_j , V'(\phi_j)\simeq 0, \langle V_{I} \rangle =0$ such that $\langle V_{I}(\phi_j) V_{I}(\phi_k)\rangle  = W \delta(\phi_j - \phi_k)$ for any $k$. Unfortunately, it is now trivial that with $V_L$ given above by ~\ref{landscapepot}, ~\ref{interactionpot}, the bubble solution of equation ~\ref{fieldsolution} does not satisfy equation ~\ref{cdlequations}. The only exception is the case when the landscape would contain one or two vacua only, with a potential of the form $V(\phi)\simeq \frac{\mu^2}{2} ( \phi - \phi_j)^2. $ . The landscape does not contain two vacua, it contains about $10^{600}$ of them and bubbles are not solutions to such a complex potential.

\section{The Wheeler-de Witt Equation}

Our starting point in investigating the state of the Universe is based on the 
canonical approach to quantum gravity. Necessarily we need a gravitational component
to our model which we will take to be general relativity. On any distance scale larger than the Planck scale, gravitation is controlled by the Einstein-Hilbert action
\beq
I_{grav} = \frac{1}{16\pi G} \int d^4x \ \sqrt{-g}\ (R - 2\Lambda)
\eeq
in which $G$ is Newton's constant, $R$ is the Ricci scalar formed from the spacetime metric
$g_{ab}$ of signature $(-\,+++)$ and $\Lambda$ is the cosmological constant. With our conventions
$\Lambda > 0$ corresponds to de Sitter space rather than anti-de Sitter space. In addition to the gravitational field, we assume that there is a scalar field 
$\phi$. 
The action for a single scalar field is taken to be
\beq
I_{scalar}= \int d^4x\ \sqrt{-g}\ \biggl(-\frac{1}{2}g^{ab}\partial_a\phi \partial_b\phi - V(\phi)\biggr) 
\eeq
where the scalar field has a potential $V(\phi)$.

As many have done before us, we investigate quantum cosmology in a small minisuperspace, defined by the scale factor of three geometries $a$ and the landscape field $\phi$.
We will take the universe to be described by Friedmann-Robertson-Walker model 
with $k=0,\pm1$ depending on whether the universe is flat $k=0$, closed $k=1$, or hyperbolic $k=-1$.
Let $\gamma_{ij}$ be the metric on these constant curvature spaces, so that its Ricci scalar
$R(\gamma)=6k.$ We take the spacetime metric to be
\beq
ds^2 = - N^2dt^2 + a^2(t)\gamma_{ij}dx^idx^j,
\eeq
where $N$ is  the shift and $a(t)$ is the scale factor of the Universe. $\phi$ we will take to be only a function of time and not of 
position. Classically, one finds that the total action under then reduces to
\beq
I =  \int \sqrt{\det\gamma}~d^3x~ dt~ Na^3\biggl(\frac{1}{16\pi G}\biggl(\frac{6\ddot a}{aN^2}
+\frac{6\dot a^2}{a^2N^2}+\frac{6k}{a^2}-6\frac{\dot a \dot N}{aN^3}-2\Lambda\biggr) +
\frac{\dot\phi^2}{2N^2} - V(\phi)\biggr). \label{eq:lagrange}
\eeq
Henceforth, we will use units such that $16\pi G=1$. Our task is now to construct a canonical version 
this system. In so doing, we will find a system with  first-class constraints, the corresponding  
gauge invariance can be fixed by imposing some suitable gauge condition on the shift, $N$. In order to find
the momenta $\pi_N, \pi_a$ and $\pi_\phi$ canonically conjugate to the variables parametrizing the minisuperspace, $N, a$ and $\phi$
respectively, we must first carry out a partial integration on the action in (\ref{eq:lagrange})~
so as to eliminate the second derivatives of $a$. We then find the momenta
\beq
\pi_N=0\ \ \ \ \pi_a= -\frac{6a\dot a}{N} \ \ \ \ \pi_\phi=\frac{a^3\dot\phi}{N}.
\eeq
The vanishing of $\pi_N$ is a primary constraint, which in turn leads to the secondary 
constraint
\beq
{\mathcal H} = -\frac{\pi_a^2}{12a} + \frac{\pi_\phi^2}{2 a^3} - 3ka +\Lambda a^3 + V(\phi)a^3,
\eeq
so that $\mathcal H$ is the minisuperspace analogue of the Hamiltonian constraint of
general relativity. The Hamiltonian $H$ for this system is therefore
\beq 
H= \int d^3x\ \  N\mathcal H \ + \lambda \pi_N
\eeq
where now $\lambda$ is a Lagrange multiplier. Given $H$, it is easy to check that this Hamiltonian 
yields the same equations of motion as the Lagrangian. $N$ is a gauge degree of freedom, and
it can be fixed in a way that is convenient for further calculations.  Fixing $N=a$
yields the metric in a conformally flat form, so that $t$ would be \lq\lq conformal \rq\rq time. 
Fixing $N=1$ makes
$t$ equal to proper time for observers at rest in the $x^i$ co-ordinate system. 
We will use the $N=1$ gauge. 

To make the transition to quantum mechanics one regards the Hamiltonian constraint as an operator
acting on a state $\Psi(a,\phi)$.  The
Hamiltonian constraint viewed as an operator must annihilate this state, $\Psi[a,\phi]$, often called the wavefunctional of the universe. 

The operator versions of the momenta conjugate to $a$ and $\phi$ are given by
\beq
\pi_a = -i \frac{\partial}{\partial a} \ \ \ \ \pi_\phi = -i \frac{\partial}{\partial \phi}
\eeq
so that the pairs $a,\pi_a$ and $\pi_\phi,\phi$ both obey the Heisenberg algebra. All of the remaining pairs of variables are taken to commute with each other. 

The only place then that causes any difficulty in realizing an operator version of $\mathcal H$ is
in the term $\pi_a^2 a^{-1}$. There is an operator ordering problem in that term. The most general 
version of this operator is 
\beq a^\alpha \pi_a a^\beta \pi_a a^{-(1+\alpha+\beta)} \eeq
for so constants $\alpha$ and $\beta$. We first ask if there is some way of
fixing what these constants can be. 
We choose an operator ordering such that the kinetic energy part of the Hamiltonian constraint
is the Laplace-Beltrami operator on some manifold. This may seem arbitrary but the reason for adopting this is because one knows that the classical vacuum of canonical general relativity can be described as being related to geodesic motion in superspace, the space of all spatial metrics modulo diffeomorphisms. 
Requiring this to be the case leads to the choice $\alpha=-2$ and $\beta=1$. The corresponding  $2$-dimensional Lorentz-signatured metric on the minisuperspace is then
\beq
ds^2 = -12a da^2 + 2a^3d\phi^2.
\eeq
Since $a \ge 0$, we see the metric on the space of all field configurations is conformal to Rindler space. 

Let us look at the classical solutions of these equations with $V(\phi)=0$.  
The classical Einstein equations are solved, for expanding universes, by
\beq
 a(t) =   ({\frac{3}{2}})^{1/6} t^{1/3}\ \ \ \ \phi(t)= ({\frac{2}{3}})^{1/2}\ln t. 
\eeq
The universe then starts to expand with $a=0$ and $\phi\rightarrow -\infty$ at  $t=0$. 
The curves in minisuperspace are given therefore given by 
\beq a=(\frac{3}{2})^{1/6}e^{(\phi-\phi_0)/\sqrt{6})}
\eeq
This is a null curve in the minisuperspace. Furthermore, it is easy to see that
it is in fact a geodesic in this metric with $t$ being an affine parameter.

Now that we have the Wheeler-deWitt equation in an acceptable form,  we need to check that the kinetic energy operator is self-adjoint given the minisuperspace metric and the induced measure on the configuration space $(a,\phi)$. From inspection of the metric on minisuperpsace, we conclude that
the measure is 
\beq \int da\ d\phi \ a^2. \label{eq:norm}\eeq
The Laplace-Beltrami operator is self- adjoint with this measure. 
It is probably possible to make other choices to resolve the operator ordering ambiguity, but this is a particularly simple one with nice properties and so we will use it from now onwards.

\section{Klein-Gordon cosmology}

As a simple warmup exercise, lets looks at the wavefunction of the universe for the case of a vanishing potential for the scalar field, with additionally both $\Lambda=0$ and $k=0$.
The Wheeler-de Witt equation is then 
\beq
\Biggl(-\frac{1}{6a}\frac{\partial^2}{\partial a^2} -\frac{1}{6a^2}\frac{\partial}{\partial a}
+\frac{1}{a^3}\frac{\partial^2}{\partial\phi^2}\Biggr)\Psi=0.\eeq
Upon the substitution $a=\sqrt{6}e^x$, the
Wheeler-de Witt equation becomes just the ordinary two-dimensional wave equation
\beq
(-\frac{\partial^2}{\partial x^2}+\frac{\partial^2}{\partial \phi^2})\Psi = 0.
\eeq
The solutions of this equation are arbitrary functions of $x \pm \phi$. Amidst such a richness
of possible solutions, we should ask which solutions are of physical interest. Firstly, we suppose that the solutions must be normalizable. Since $a$ and hence $x$ is being treated as the analog of time, normalizable means that
the integral of the square modulus of $\Psi$ over $\phi$ is finite.
Thus we expect
\beq
\int d\phi \Psi^\ast\, \Psi < \infty
\eeq
If so, one can then compute the  expectation values of 
various quantities. A second requirement is that we can relate wavefunctions to measurable and observable quantities. One would like, for example, to find a wave function that describes our universe. 
We know that in the semi-classical limit, a wavefunction should have constant phase on solutions of the classical equations. Looking at the classical solutions found in the previous section, we immediately can conclude that $x - \phi$ is constant for solutions where the universe is expanding. We are then confronted with the obvious problem that the solutions of the Wheeler-de Witt equation do not explicitly involve time. Whilst that is true, it is not the case that
time is not part of the description. If we ask about an observation on our universe, we are going to make measurements of physically measurable quantities like the size of the universe, the velocity with which it is expanding or the values of the scalar field. We are not going to make any measurement of the value of the time co-ordinate. If we ask for a typical wavefunction that describes something like our universe, 
we are going to say that it is peaked around the values of $\phi_i$ that we measure when the scale factor is 
$a_i$ or equivalently when $x=x_i=\sqrt{6}\ln a_i$. A suitable wavefunction might then be, for fixed $a_i$,
\beq
\Psi \sim e^{-{\lambda/2}(\phi-\phi_i)^2}
\eeq
where $\lambda$ is some parameter that reflects the precision to which we have measured $\phi$.
Since we know that this was measured when the Universe has scale-factor $a_i$ and the wavefunction for an expanding universe is given by an arbitrary function of $x-\phi$, we find the wavefunction to be in general
a superposition of plane-wave type solutions that give a gaussian distribution. Using methods familiar from quantum mechanics, we conclude that such a wavefunction is
\beq
\Psi_(x,\phi) = \lambda^{1/4}\pi^{-{1/4}}e^{{-{\lambda/2}(x-x_i-\phi+\phi_i)^2}}
\eeq

Suppose now that
one asks for the probability of measuring $\phi$ and getting the answer $\phi_f$ when the universe
is measured to have a scale factor $a_f$ and so $x_f=\sqrt{6}\ln a_f$. 
The amplitude $A_{fi}$ for this observation 
is then given by the overlap integral
\beq
A_{fi} = \int \ d\phi \Psi^*(x_f,\phi_f)\ \Psi(x_i,\phi_i).
\eeq
The corresponding probability distribution for measuring $\phi_f$ is then peaked around the classical solution, rather as one might have expected.

\section{Periodic Potentials}

It is hard to extend these models using general potentials, but we can easily gain some insight
into the question by looking at how one treats  some simplified models of the landscape potential. 
It is straightforward to model a \lq periodic landscape' by a collection of $\delta$-function potentials,
that is to say that
\beq
V(\phi) = \sum_n g_n \delta(\phi-\phi_n). 
\label{eq:mdf}
\eeq

To start this section, we will look at a simple example, where the potential is a single delta function.
Let 
\beq V(\phi) = g \delta(\phi).
\eeq
If $g>0$, then we will just find wavelike solutions as before. However if $g<0$, then one might expect to find a bound state with the wavefunction concentrated at $\phi=0$. The Wheeler-de Witt equation becomes under these circumstances
\beq
\Biggl(-\frac{1}{6a}\frac{\partial^2}{\partial a^2} -\frac{1}{6a^2}\frac{\partial}{\partial a}
+\frac{1}{a^3}\frac{\partial^2}{\partial\phi^2}+ 2ga^3\delta(\phi)\Biggr)\Psi=0.\eeq
Solutions of this equation are of the same form as when the potential vanishes as long as one is away from $\phi=0$. Suppose the solution is 
\beq \Psi = a^{\sqrt{6}ik}(\alpha_\pm e^{ik\phi} + \beta_\pm e^{-ik\phi})
\eeq
with the minus sign being for $\phi<0$ and the plus sign for $\phi>0$. The wavefunction is 
continuous at $\phi=0$ but has a discontinuity in its derivative with respect to $\phi$ such that
\beq
{\rm Lt}_{\epsilon\rightarrow 0} \Bigl(\frac{\partial\phi(\epsilon)}{\partial\phi} - \frac{\partial\phi(-\epsilon)}{\partial\phi}\Bigr) = 2ga^6\phi(0).
\eeq

This requires
\beq \begin{pmatrix} \alpha_+\\ \beta_+ \end{pmatrix} = \begin{pmatrix} 1-iX & -iX \\ iX & 1+iX \end{pmatrix}
\begin{pmatrix} \alpha_- \\ \beta_- \end{pmatrix} \eeq
where $X=ga^6/k$.
For $g>0$, the wavefunctions are not localized at $\phi=0$. In all cases, they are similar to those described in the previous section except that they are necessarily superpositions of universes in which,
according to the interpretation given in the previous section, increasing values of $a$ are now 
correlated with both increasing $\phi$ and decreasing $\phi$. 

If $g<0$, then one can find wavefunctions that are concentrated at $\phi=0$, being exponentially suppressed 
away from $\phi=0$. These solutions are analogous to those found in simple quantum mechanical problems where 
an attractive delta function potential in one dimension has a single bound state. However, for precisely this reason the wavefunctions do not have classical behaviour as surfaces of constant phase do not resemble 
our universe. 

Suppose instead we considered a periodic potential given the potential in equation (\ref{eq:mdf}).
Between $\phi=n\Delta$ and $\phi=(n+1)\Delta$, the solutions of the Wheeler-deWitt equation have the same functional form as before but now there are discontinuities in the gradient of the wavefunction
at each of the delta-functions. Solutions are then of the form
\beq
\Psi_n(a,\phi) = (\alpha_n(k) e^{ik(\phi-n\Delta)} + \beta_n(k) e^{-ik(\phi-n\Delta)})a^{\sqrt{6}ik}.
\eeq
Matching at the locations of the delta-functions then leads to a recursion relation for the
coefficients $\alpha_n(k)$ and $\beta_n(k)$. 
\beq
\begin{pmatrix}  \alpha_n(k)\\ \beta_n(k) \end{pmatrix} = \begin{pmatrix} (1+X)e^{ik\Delta} 
& Xe^{-ik\Delta} \\ -Xe^{ik\Delta} & (1-X)e^{-ik\Delta} \end{pmatrix}
\begin{pmatrix} \alpha_{n-1}(k) \\ \beta_{n-1}(k) \end{pmatrix} \eeq
where now  $X=-\frac{iga^6}{k}$.

The transfer matrix so defined has unit determinant, thus its eigenvalues are $\lambda$ and $\frac{1}{\lambda}$. If $\lambda$ is real, then  the wavefunction does not have any solutions that
are bounded as $\phi \rightarrow \pm\infty$. The roots will be complex conjugate pairs of unit modulus
if $Y^2<1$ where
\beq
Y=\cos k\Delta + \frac{ga^6}{k}\sin k\Delta.
\eeq

This latter situation gives rise to bands of allowed energy, just like a one-dimensional solid. The wavefunction is not localized near any of the minima of this potential. It is of the Bloch wave form, extended over the whole potential. One might worry that these bands
would not persist as $a$ gets large. The boundary of the bands is at 
\beq
\cos k\Delta = \pm \frac{1-Y^2}{1+Y^2}.
\eeq.
Since $-1 \leq \frac{1-Y^2}{1+Y^2} \geq 1$, there is always a solution to the previous equation. 
The wavefunction of such universes seems to be completely unlike any behavior we expect for a realistic
universe model. We will denote these solutions for the wavefunction in a periodic potential with N-sites by $\Psi_p[a,\phi]$

\section{Non-periodic Potentials}

We will now break the periodicity in equation ~\ref{eq:mdf} by changing the coupling constant $g_n$ for one site, to $\tilde{g} \neq g_n$

\beq
V(\phi) = \tilde{g} \delta(\phi) + \sum_n g_n \delta(\phi+ n l). 
\label{eq:nonp}
\eeq

With this potential the Wheeler DeWitt equation becomes
\beq
\Biggl(\frac{\partial^2}{\partial a^2} +\frac{1}{a}\frac{\partial}{\partial a} \Biggr)\Psi[a,\phi] = 
\Biggl(\frac{6}{a^2}\frac{\partial^2}{\partial\phi^2} + 12 \frac{a^4}{h^2} V(\phi) \Biggr)\Psi[a,\phi].
\label{eq:wdwnonp}
\eeq

We have rewritten this equation with the gravitational part on the left hand side, and the field Hamiltonian in the right hand side such that, $H_a \Psi[a,\phi]= - H_{\phi} \Psi[a,\phi]= E_a \Psi[a,\phi]$, where $E_a$ is the energy eigenvalue of the right hand side, which depends on $a$. We can solve the right hand side of equation ~\ref{eq:wdwnonp} 

\beq
\Biggl(\frac{6}{a^2}\frac{\partial^2}{\partial\phi^2}+ 12 \frac{a^4}{h^2}[ \tilde{g} \delta(\phi) + \sum_n g_{n} \delta (\phi + n l) ] \Biggr)\Psi[a,\phi]= E_{a}\Psi[a,\phi].
\label{nonpfield}
\eeq

By replacing the ansatz $E_a = E^0_{a} + \delta E_a$ and $\Psi[a,\phi] = \Psi_0[a] e^{\pm \kappa_a \phi} \Psi_p $ in equation ~\ref{nonpfield}, where $\Psi_p$ and $E^0_{a}$ are wavefunction and eigenvalue solutions of the periodic potential of the previous section, or equivalently solutions to equation ~\ref{nonpfield} for $\tilde{g} = 0$, we obtain

\beq
(\frac{6}{a^2} \kappa^2_{a}  -\delta E_a )e^{\pm \kappa_a \phi} = -( 12\frac{a^4}{h^2}\tilde{g} \delta(\phi) \pm \kappa_a \frac{12 \Psi'_p}{a^2 \Psi_p} \Biggr)e^{\pm \kappa_a \phi} . 
\label{field}
\eeq

Here $\Psi_0(a)$ is the part of the wavefunction that does not depend on the field $\phi$, which solves the left hand side of equation ~\ref{eq:wdwnonp}, $- H_a \Psi_{0}[a] = ( \frac{\partial^2}{\partial a^2} +\frac{1}{a}\frac{\partial}{\partial a})\Psi_0 = E_a \Psi_{0}[a]$. The solution to this equation is a Bessel function, $\Psi_0 [a]\simeq a^{1/2} J_{1/6} (\kappa_a)$.

Integrating equation ~\ref{field} around $\phi=0$, we get
\beq
\kappa_a = - \tilde{g} a^6.
\eeq
and 
\beq
\delta E_a = \frac{\kappa_a^{2}}{2 a^2} = \tilde{g}^2 a^{10} .
\eeq

Finally, putting everything together we have: $\Psi[a, \phi] = N J_{1/6}(\kappa_a)e^{\kappa_a \phi} \Psi_{p}[a, \phi]$, and energy $E_a = E^{0}_a + \frac{h^2 \kappa_a^{2}}{2 a^2}$. The emergence of localization can be seen even with one disordered site. 
\subsection{Random Potentials}
If we were to make all the coupling constants in the potential of equation ~\ref{eq:nonp} be drawn randomly from an interval, then we would have an example closely related to the landscape potential of \cite{denef} in which $\langle V(\phi \rangle =0$ and $\langle V(\phi_i) V(\phi_j)\rangle = \Gamma \delta (\phi_i - \phi_j)$ with $\Gamma$ the disorder strength. Solutions of the WDW equation for the wavefunction of the universe propagating on the string theory landscape were studied in ~\cite{laura}. We can see how this solution can be obtained from the previous example, by increasing the number of disordered vacua in the above case of a non-periodic potential, one by one and finding the solutions iteratively. These types of potentials were first studied by Anderson \cite{andersonlocal}. More sophisticated methods were developed later \cite{efetov, altland}. Through a redefinition of the variables $a,\phi$ to $\alpha =ln(a), x =a^3 \phi$, one obtains from the WDW equation $\Psi[x,\alpha] = =\sum_j \Psi_j [x,\alpha]= \sum_j \psi_j [x]F_J [\alpha]$ where

\beq
H(x) \psi_j(x) = \epsilon \psi_j (x).
\eeq
and

\beq
- \frac{\partial^2}{\partial \alpha ^2} F_j (\alpha) = \epsilon_j F_j (\alpha).
\eeq
with $\epsilon_j$ having an imaginary part proportional to the disorder strength of the landscape $\Gamma$, explained below. The solution for the wavefunctions are the Anderson localized ones
\beq
\Psi_j [\alpha, x] \simeq \frac{1}{\epsilon^{1/4} l^{1/2}_j} e^{\pm i \sqrt{\epsilon_j}\alpha - \frac{x - x_j}{2 l_j}}.
\label{anderson}
\eeq
where the localization length $l_j$ around site $j$ is given by $l_j^{-2} \simeq Im[\epsilon_j]$. Thus for a disordered large dimension potential, such as the landscape, localization always occurs for all the wavefunctions with energies below the disorder strength. Since the disorder of the landscape is of order the string scale then all wavefunctions will localize. We can now proceed to understand the physics behind the landscape and bubble solutions.

\subsection{Comparing Landscape Solutions to Bubble Solutions}

The lattice type potentials with a random  distribution of vacua locations and of energies, known as disordered potentials, are well known in condensed matter \cite{efetov,altland} since many systems such as spin glass and quantum dots fall into this category. In cosmology each solution for the wavefunction on the landscape potential gives rise to a universe with the geometry determined by the well where the wavefunction localizes, via Einstein equations. That is, each wavefunction solution to the landscape potential in the family of solutions gives rise to its own geometry which itself becomes a random variable. In order to incorporate the randomness of geometries, and the consistency of quantum wavefunction solutions with Einstein equations, we had to treat this problem within canonical quantum gravity, with geometries and fields becoming quantum operators rather then a point in spacetime. The  Einstein equations and quantum mechanics equations are recovered \cite{kiefer,hawking} in the semiclassical limit, via the identification $\nabla S \dot \nabla = \frac{\partial}{\partial t}$ with $S$ the Euclidean action.

Cosmological solutions of the wavefunction for the case of the string theory landscape potential were studied and discussed in detail in \cite{laura}. A disordered potential gives rise to the Anderson localization phenomenon, where multiple scattering of the wavefunction over many wells and barriers of the potential leads to destructive interference and thus localization of the wavefunction on one of the wells. Of course tunneling from a potential well to the next one is explicitly included in the transport of the wavefunction which undergoes multiple scattering.  Anderson localization is an inherently quantum mechanical phenomenon since it is based on quantum mechanical interference and thus can not be studied by quantum field theory methods. CDL instantons or bubble profiles similar to them, can also be studied by quantum mechanics methods and that solution is very well known. Therefore we can compare the two solutions, the bubble and the landscape solution for the wavefunction.  

Instanton solutions for a double well potential, have been extensively studied within the quantum cosmology framework and in the semiclassical limit, see for example ~]\cite{vilenkintunnel, hh}. Although the particular semiclassical solution of the double well potential depends on the choice of the boundary conditions, these solutions share the feature of exponentially suppressed density of states below, independent of the choice of boundary conditions.  
The energy shift due to tunneling in a double well pair is $\delta \epsilon \simeq \Gamma$ where $\Gamma \simeq e^{- S_E}$ is the nucleation rate of CDL bubbles with a Euclidean action $S_E$. The density of states $\rho \simeq \Gamma$ and the two point function for this case is 

\begin{equation}
G(\phi_j , \phi_{j \pm 1} ) \simeq \sinh[\Gamma \tau /2] \rightarrow_{\tau->\infty}  e^{-\Gamma \tau/2}  \,
\label{bubblegreen}
\end{equation}

As shown in \cite{laura} for the landscape potential $V_{L}(\phi)$ energies are shifted in the complex plane, $\epsilon_j = \epsilon^{0}_{j} - \delta \epsilon_j  - i \gamma_j $ where $\gamma_j = \frac{2\pi W}{\epsilon_j} \simeq \frac{1}{l_j}^2$. The solution of the wavefunction for this potential $V_{L}(\phi)$ found in ~\cite{laura} is $\Psi(\phi) \simeq \frac{1}{l_j} e^{-\frac{(\phi -\phi_j)}{2l_j}}$ , the Anderson localized solution. Here $l_j \simeq \sqrt{1/\gamma_j}$ is the localization length. The averaged density of states for the landscape potential is given by the imaginary part of he advanced Green's function: $Im G_{A}(\phi_j) \simeq \frac{\gamma_j}{(\epsilon -\epsilon_j)^2 - \gamma_j^2}$ with poles at $|\epsilon| = |\epsilon_j - \delta\epsilon_j - i\gamma_j|$ which yields a density of states 

\begin{equation}
\rho(\epsilon) \simeq \frac{1}{|\epsilon| + 1/l^2}  \,
\label{landscapedensity}
\end{equation}
in contrast to the exponentially suppressed distribution of the dilute gas of CDL bubbles. {\it Note $\rho(\epsilon)$ falls off as a power law with the energy. Thus unlike false vacuum decay, high energy states have a nonnegligible probability for the landscape potential. The Fourier transform of the Greens function that solves the quantum equation is $G(\Psi(\phi) , \Psi(\phi_j) ) \simeq e^{- (\phi -\phi_j)/l}$, (similar to spin glass, ~\cite{spinglass})}.

We can try and shed some light into the physical reason behind the mathematical inconsistency of the two solutions.
In the case of the landscape, the potential $V_{L}(\phi)$ is roughly of the white-noise type \cite{landscape}. Solutions to such potentials are known as Anderson localization \cite{andersonlocal} and they are physically different from the bubble ones. It should be emphasized that the dynamics of the wavefunction on the landscape potential $V_{L}$ is based on N-vacua physics, where tunneling from a vacuum state to its neighbor is only part of the dynamics and transport. Besides tunneling, a very important ingredient that is missing from the treatment of \cite{susskind} but which is responsible for the transports of the wavefunction through many vacua in $V_L$, is the quantum effect of constructive and destructive interference among the phases of the field resulting from multiple scattering of the field through many of the N-vacua of the landscape. Multiple scattering is present and is the dominant mechanism, even in the nearest neighbor approximation. Physically this is the reason why our universe $\Psi_{us}$ is influenced not only by its nearest neighbor but by many vacua, in fact by the whole structure of vacua distributions on the landscape. It is straightforward to see that even with tunneling to the nearest neighbor as the only mechanism for the field, the field will transport and scan the whole structure of the landscape since each 'true' vacuum neighbor to us will have its own nearest neighbor, which will have its own nearest neighbor and so on. In short all vacua neighbors of neighbors are unstable to tunneling. In principle the field's transport scans the landscape structure, finally being localized in one vacuum state induced by disorder. The landscape problem can not be reduced to a 2-body problem and it can not be solved as a tunneling event among two vacuum states alone. A key contribution to localization of $\Psi$ on some vacuum state, and therefore to the production of a universe,  comes from quantum interference from multiple scattering across the N-vacua disordered landscape.

\section{Conclusion}
We have shown that bubble solutions produced by eternal inflation are mathematically inconsistent with the landscape potential. The reason is that due to interference induced from multiple scattering and multiple tunneling, the wavefunction scans a larger part of the landscape than its nearest neighbor. The disordered landscape leads to destructive interference and localization. The vacua where the wavefunction localize can not be apriori determined, therefore the vacua can not be anthropically selected.

We argue that the best available formalism for treating both, the field and the three-geometries, as independent random variables, when addressing fundamental questions like the origin of the universe, is quantum cosmology. We provide solution for the wavefunction of the universe from the landscape. The landscape solutions are Anderson localized. We compare the landscape solutions to the bubble ones and reveal that quantum interference is the physical reasons for the mathematical inconsistency between them.

Besides the mathematical inconsistency of field solutions, another concern is the fact that vacuum energy of each vacua site is a randomly chosen variable, therefore the background de Sitter space arising from 'false vacuum decay' becomes a random variable itself. Eternal inflation on the landscape leads to a situation where the geometries of all bubbles are independent and they are all embedded in a randomly varying and randomly splitting background. With gravity included, the situation becomes worse. The landscape potential gives rise to a totally different spacetime from the eternal inflation one. Anthropic arguments for choosing the right pair of neighbors on this complex landscape further complicates the validity of this picture since the probability distribution of solutions is only polynomially suppressed at high energy vacua, (i.e. transitions to the high energy vacuum states have a nonnegligible likelihood). Besides, as shown in ~\cite{fred} anthropic selection, such as the Weinberg's arguments ~\cite{weinberg} the authors of ~\cite{susskind} appeal to, actually favor a universe where observes suffer a cosmic heat death.

We did not include decoherence in this treatment. In \cite{susskind} fluctuations were considered at the level of structure formation. In reality including fluctuations prior to the emergence of spacetime from the wavefunction, plays a crucial role in the resulting geometry of spacetimes and the probability distribution of solutions, especially in terms of the randomly varying background DS geometry \cite{holman}.

In closing, recently a lot of work is focused on observational signatures of bubble collisions from eternal inflation in the framework of the landscape. But in fact we have an abundance of observational evidence that the landscape is not populated by eternal inflation bubbles. Every material known in nature can be describes as a large landscape of atoms with N potential well sites, just like the moduli landscape, and is studied within a quantum field framework. If it is correct that all these structures can be reduced to, and equivalently be described by, a single or double well potential,( that is reduced to one or two atoms), then the only existing matter in the universe should be bubbles. We know we do not live in a universe where stars,structure and all matter around us is made of bubbles, therefore we have sufficient observational evidence that the eternal inflation in the framework of the landscape is incorrect, it is a nonsequitur. Physically the reason that reducing the large landscape  to a pair of wells is incorrect, lies on the fact that the rich quantum dynamics contained in the wavefunction scattering through an intricate structure such as the landscape, should not and can not be replaced with a bounce solution in a double well, as we just demonstrated here. 

We have shown in this letter that claims for eternal inflation populating the landscape are incorrect. Eternal inflation has no known relation to the landscape. The latter has further implication for the string theory landscape. we recently showed that eternal inflation can not even be eternal thus it can not produce a multiverse ~\cite{noei}. On the other hand we have shown here that the landscape is independent of eternal inflation, and it does give rise to a family of solutions for the wavefunction of the universe and through the wavefunction it is closely related to the many worlds interpretation of quantum mechanics.
Therefore any observational constraints placed on bubble collision scenarios should not be interpreted as a test of the string theory landscape.

{\it Acknowledgements}
 LMH would like to thank DAMTP,University of Cambridge, for their hospitality where this work was done. LMH acknowledges the DOE support of grant DE-FG02-06ER1418 and the Bahnson fund.


\end{document}